\documentclass[showpacs,twocolumn,prl]{revtex4}
\usepackage{graphicx, amsmath, amsthm, amssymb}
\newcommand{\be}{\begin{equation}}
\newcommand{\ee}{\end{equation}}
\newcommand{\bea}{\begin{eqnarray}}
\newcommand{\eea}{\end{eqnarray}}
\newcommand{\bwt}{\begin{widetext}}
\newcommand{\ewt}{\end{widetext}}
\newcommand{\ra}{\rangle}
\newcommand{\la}{\langle}
\newcommand{\up}{\uparrow}
\newcommand{\dn}{\downarrow}

\newcommand{\dg}{\dagger}

\newcommand{\cD}{{\cal D}}
\newcommand{\cH}{{\cal H}}
\newcommand{\cZ}{{\cal Z}}
\newcommand{\cS}{{\cal S}}
\newcommand{\bsco}{Bi$_{2}$Sr$_{2}$CaCu$_{2}$O$_{8+x}$ }
\newcommand{\bpsco}{Bi$_{2-y}$Pb$_{y}$Sr$_{2}$CaCu$_{2}$O$_{8+x}$ }
\begin{document}
\title{Phase Fluctuations in High Temperature Superconductors}
\author{Wonkee Kim$^{1}$, Yan Chen$^{2}$, and C. S. Ting$^{1}$}
\affiliation{
$^1$ Texas Center for Superconductivity and Department of Physics, 
University of Houston, Houston, Texas 77204
\\
$^2$ Department of Physics and Lab of Advanced Materials, Fudan University, 
Shanghai, China
}

\begin{abstract}
Within the phase fluctuation picture for the pseudogap state of a high-$T_{c}$
superconductor, we incorporate the phase fluctuations generated by the classical
XY model with the Bogoliubov-de Gennes formalism utilizing a field-theoretical 
method.
This picture delineates the inhomogeneous characteristics of
local order parameters observed in high-$T_{c}$ superconductors
above $T_{c}$.
We also compute the local density of states near a non-magnetic impurity 
with a strong scattering potential. The resonance peak smoothly evolves
as temperature increases through $T_{c}$ without showing any sudden broadening,
which is consistent with recent experimental findings. 
\end{abstract}


\date{\today}
\maketitle

One of the most defining features of high-T$_{c}$ superconductors,  
the pseudogap state, has drawn very intense attention in recent years\cite{timusk}.
Various theoretical models have attempted to explain this state; they generally fall
into two categories. One is the pre-formed
pair model, or phase fluctuation scenario\cite{emery,sharapov,eckl,kim,benfatto,dubi}.
In this picture, the Cooper pairs continue to exist above $T_{c}$, in the pseudogap
state below $T^{*}$ without phase coherence, where the
superfluid stiffness is zero. A second category is the competing gap scenario, where,
for example,  
a density-wave gap coexists with the superconducting gap\cite{chakravarty,zhu,sachdev,kivelson}.
These distinctions can be fairly subtle, and the nature of this phase continues
to attract growing interest.
Angle-resolved photo-emission spectroscopy experiments\cite{ding} have supported 
the phase fluctuation scenario since the pseudogap seems to evolve into 
the superconducting gap as the temperature changes through
$T_{c}$. Further support can be found in the Nernst experiments\cite{xu,wang},
where a large Nernst signal induced by vortex motions is observed above $T_{c}$ 
in hole-doped cuprates. This result has been interpreted as evidence for phase
fluctuations. 
In the pseudogap state, from the viewpoint of the phase fluctuation scenario,
one can conceive a spatially dependent order parameter 
$\Delta({\bf r})=|\Delta|e^{i\theta({\bf r})}$
at a location ${\bf r}$\cite{thermal_ave}.
In fact, an intrinsic inhomogeneity\cite{pan0,lang} of local order
parameters has been observed even at a low temperature.
In a two-dimensional theory\cite{sharapov,eckl,kim,benfatto,dubi}, 
the transition into the superconducting state
at $T_{c}$ is not BCS-like. Instead, it is the Kosterlitz-Thouless (KT) transition\cite{kt}.
Consequently, we identify $T_{KT}=T_{c}$, which depends on the superfluid stiffness.
Denoting $T_{MF}$ as the temperature when
Cooper pairs preform, we also define $T_{MF}=T^{*}$, which is determined by the strength
of the pairing potential.

Recently, spatial variations of the order parameter were visualized in topographic
images\cite{gomes,chatterjee}. Gomes {\it et al.}\cite{gomes} have examined statistically
the evolution of the order parameter in atomic scale of \bsco for various doping cases
with increasing $T$ from a low $T$ through $T_{c}$ to a high $T$.
They illustrated the distribution $D$ as well as the probability $P$ of 
local order parameters.
The distribution $D(\Delta)$ is an ordered list of local order parameters, 
which describes how many sites have
the order parameter less than $\Delta$. Thus $D(\Delta+\epsilon)-D(\Delta)$ gives the number of
sites with the order parameter between $\Delta+\epsilon$ and $\Delta$. Since $P(\Delta)$
is the probability to find a local order parameter of $\Delta$, 
$P(\Delta)\propto \partial_{\omega} D(\omega)|_{\Delta}$.
On the other hand, Chatterjee {\it et al.}\cite{chatterjee} observed that the impurity resonance
survives for $T > T_{c}$ in the study of the scanning tunneling microscopy (STM) for \bpsco.
The STM study\cite{pan} for high-$T_{c}$ superconductors is performed usually at a low 
$T\ll T_{c}$. Since $T_{c}$ of this compound is $15K$, they carried out the
$T$ dependence of the impurity resonance through $T_{c}$.
In this paper, we wish to demonstrate that these recent findings\cite{gomes,chatterjee} are 
understandable, at least qualitatively, within the framework of the phase fluctuation scenario.

In the theoretical formulation, we separate the order parameter into the mean-field value and 
the phase fluctuation part using the field-theoretical approach. 
The mean-field value is determined by the 
the Bogoliubov-de Gennes(BdG) formalism while phase fluctuations will be 
generated by the classical XY model.
Let us start with
the Hamiltonian of a $d$-wave superconductor
with a definition of 
${\hat\psi}^{\dg}_{i}=(C^{\dg}_{i\up}, C_{i\dn})$ and the Pauli matrices
${\hat\tau}_{3}$ and ${\hat\tau}_{\pm}=({\hat\tau}_{1}\pm i{\hat\tau}_{2})/2$,
\bea
\cH =&-&\sum_{\la i,j\ra} t_{i,j}{\hat\psi}^{\dg}_{i}{\hat\tau}_{3}{\hat\psi}_{j}
+\sum_{i}(U_{i}-\mu){\hat\psi}^{\dg}_{i}{\hat\tau}_{3}{\hat\psi}_{i}
\nonumber\\
&-&\sum_{i,j}V_{i,j}{\hat\psi}^{\dg}_{i}{\hat\tau}_{+}
{\hat\psi}_{j}{\hat\psi}^{\dg}_{j}{\hat\tau}_{-}{\hat\psi}_{i}
\eea
where $t_{i,j}$ is the hopping amplitude,
$U_{i}$ is an impurity potential at a site $i$, 
$\mu$ is the chemical potential, $V_{i,j}$ is a pairing potential,
and ${\bar\sigma} = -\sigma$ with $\sigma = \up,\;\dn$.
The symbol $\la i,j\ra$ means a sum over nearest neighbor pairs.
Since no magnetic field is included, $t_{i,j} = t\delta_{i+{\bf\delta},i}$,
where ${\bf\delta}=\pm{\hat x},\;\pm{\hat y}$ are the unit vectors along the $x(y)$
direction. Using a set of impurity sites ${\cal I}$, one can write
$U_{i}$ as $\sum_{l\in{\cal I}}U_{l}\delta_{l,i}$
For the nearest neighbor pairing, $V_{i,j}=V\delta_{i+{\bf\delta},j}$.
We set the lattice constant to be unity and use units such that $\hbar=k_{B}=1$.
The partition function of $\cH$ in the path integral is
$
\cZ=\int{}\cD\psi\cD\psi^{\dg}\;e^{-\cS[\psi,\psi^{\dg}]}
$
where 
$
\cS[\psi,\psi^{\dg}]=\int{}d\tau\left[\sum_{i}{\hat\psi}^{\dg}_{i}\partial_{\tau}
{\hat\psi}_{i}+\cH\right]
$ and $\cD\psi=\prod_{i}d\psi_{i}$.
The range for the integral over imaginary time $\tau$ in the action $\cS$ 
is from $0$ to $\beta=1/T$.

Introducing the Hubbard-Stratonovich transformation with auxiliary field $\phi_{ij}$
as a field-theoretical method to deal with phase fluctuations,
we obtain the partition function as follows:
$
\cZ=\int{}
\cD\phi\cD\phi^{\dg}\cD\psi\cD\psi^{\dg}\;e^{-\cS[\phi,\phi^{\dg},\psi,\psi^{\dg}]}
$,
where, for the static case: $\partial_{\tau}{\hat\psi}_{i} = 0$,
\bwt
\be
\cS[\phi,\phi^{\dg},\psi,\psi^{\dg}]=\int{}d\tau\Bigl
[\sum_{i,j}\frac{1}{V_{i,j}}|\phi_{ij}|^{2}
-\sum_{<i,j>} t_{i,j}{\hat\psi}^{\dg}_{i}{\hat\tau}_{3}{\hat\psi}_{j}
+\sum_{i}(U_{i}-\mu){\hat\psi}^{\dg}_{i}{\hat\tau}_{3}{\hat\psi}_{i}
-\sum_{i,j}\left(\phi_{ij}{\hat\psi}^{\dg}_{i}{\hat\tau}_{+}{\hat\psi}_{j}
+\phi^{\dg}_{ij}{\hat\psi}^{\dg}_{j}{\hat\tau}_{-}{\hat\psi}_{i}\right)
\Bigr].
\ee
\ewt
In the saddle point approximation, 
$\delta\cS[\phi,\phi^{\dg},\psi,\psi^{\dg}]/\delta\phi^{\dg}=0$,
the auxiliary field can be identified as the order parameter;
namely, $
\phi_{ij} = V_{i,j}{\hat\psi}^{\dg}_{j}{\hat\tau}_{-}{\hat\psi}_{i}
= \Delta_{ij}$.
Now
$
\cS[\Delta,\Delta^{\dg},\psi,\psi^{\dg}] = 
\sum_{i,j}\frac{\beta}{V_{i,j}}|\Delta_{ij}|^{2}
+\beta\cH_{BdG}[\psi,\psi^{\dg}]
$,
where the BdG Hamiltonian is
\bea
\cH_{BdG}[\psi,\psi^{\dg}]=
&-&\sum_{<i,j>} t_{i,j}{\hat\psi}^{\dg}_{i}{\hat\tau}_{3}{\hat\psi}_{j}
+\sum_{i}(U_{i}-\mu){\hat\psi}^{\dg}_{i}{\hat\tau}_{3}{\hat\psi}_{i}
\nonumber\\
&-&\sum_{i,j}\left(\Delta_{ij}{\hat\psi}^{\dg}_{i}{\hat\tau}_{+}{\hat\psi}_{j}
+h.c\right).
\eea
The effective partition function can be written as
\be
\cZ = \int{}\cD\Delta\cD\Delta^{\dg}
\exp\left[-\sum_{i,j}\frac{\beta}{V_{i,j}}|\Delta_{ij}|^{2}\right]
\cZ_{BdG}
\ee
where $\cZ_{BdG} = \mbox{Tr}\left[e^{-\beta\cH_{BdG}}\right]$
is the partition function
corresponding to the BdG Hamiltonian.
Note that $\cZ_{BdG}$ depends on $\Delta_{ij}=|\Delta_{ij}|e^{i\theta_{ij}}$,
where $\theta_{ij}$ represents phase fluctuations (see below). 
However, as a mean field approximation,
the BdG formalism does not consider phase fluctuations. 
To incorporate the fluctuations into the framework, we utilize
the two-dimensional XY model\cite{chester} with $\cH_{XY}=-J_{XY}\sum_{<i,j>}
\cos\left(\theta_{i}-\theta_{j}\right)$, where $J_{XY}$ is the coupling strength (or
the superfluid stiffness) and $\theta_{i}$ is the angle made by a classical spin
at a site $i$. 
Following Ref.\cite{chester}, we perform simulations using the Monte Carlo
method with the Metropolis algorithm.
The lattice size in our calculations is $24\times24$ with periodic boundary 
conditions.  The minimum length of phase fluctuations is set to be 2.
Starting from a random configuration, the XY model
system is cooled down to
a working $T$. It is known that $T_{TK}\approx 0.9J_{XY}$ in numerical simulations.
To benchmark our simulations, we also compute the specific heat of the 
system, which shows similar behavior to the one computed in Ref.\cite{chester}.

The mean-field value ${\bar\Delta}_{ij}$ of the order parameter is obtained 
self-consistently 
by solving the BdG equation:
\be
\sum_{j}
\left (
\begin{array}{cc}
\cH_{ij,\sigma} & {\bar\Delta}_{ij} \\
{\bar\Delta}^{*}_{ij} & -\cH^{*}_{ij,{\bar\sigma}}
\end{array}
\right )
\left (
\begin{array}{c}
u^{n}_{j,\sigma}\\
v^{n}_{j,\sigma}
\end{array}
\right) = E_{n}
\left (
\begin{array}{c}
u^{n}_{i,\sigma}\\
v^{n}_{i,\sigma}
\end{array}
\right)
\ee
where $\cH_{ij,\sigma}=-t_{ij}+(U_{i}-\mu)\delta_{i,j}$.
The bonding order parameter is evaluated as
${\bar\Delta}_{ij}=(V/4)\sum_{n}\left[u^{n}_{i\up}v^{n*}_{j\dn}+v^{n*}_{i\dn}u^{n}_{j\up}\right]
\tanh(E_{n}/2T)$. The local $d$-wave order parameter at a site $i$ is given by
${\bar\Delta}^{d}_{i}=\frac{1}{4}\sum_{j}{\bar\Delta}_{ij}\left[
\delta_{j,i+{\hat x}}+\delta_{j,i-{\hat x}}-\delta_{j,i+{\hat y}}-\delta_{j,i-{\hat y}}\right]$.
Rigorously speaking,
$\Delta_{ij}={\bar\Delta}_{ij}e^{i\theta_{ij}}$, where 
${\bar\Delta}_{ij}=|\Delta_{ij}|e^{i{\bar\theta}_{ij}}$. 
Consequently, $\Delta_{ij}=|\Delta_{ij}|\exp\left[i({\bar\theta}_{ij}+
\theta_{ij})\right]$.
The phase ${\bar\theta}_{ij}$ associated with the mean-field value
is determined within the BdG calculations. On the other hand, 
$\theta_{ij}$ is generated based on the XY model as mentioned.
Assuming the effective phase of the local $d$-wave order parameter
varies slowly over a distance of the coherence length\cite{sharapov,eckl},
it is reasonable to define $\theta_{ij}=\left[\theta_{i}+\theta_{j}\right]/2$,
which corresponds to the phase of the center of mass of a Cooper pair
in the continuum limit.

Let us now calculate the average of $\la\Delta_{ij}\ra$ over phase fluctuations.
It is straightforward to show that
$\la\Delta_{ij}\ra={\bar\Delta}_{ij}\la e^{i\theta_{ij}}\ra$, where
\be
\la e^{i\theta_{ij}}\ra=
\frac{\int{}\cD\theta\;e^{i\theta_{ij}}\cZ_{BdG}[{\bar\Delta},\theta]}
{\int{}\cD\theta\;\cZ_{BdG}[{\bar\Delta},\theta]}
\ee
with
$\cZ_{BdG} = \prod_{n}\left(1+e^{-\beta E_{n}}\right)$.
As in Refs.\cite{eckl,dubi}, $\la e^{i\theta_{ij}}\ra$ is numerically evaluated
using the Monte Carlo method.
Generating a phase configuration $\{\theta_{ij}\}$
at $T$ and diagonalizing the BdG Hamiltonian with $\Delta_{ij}$, we obtain
$E_{n}$ corresponding to the given configuration. In the averaging process,
the total number of
phase configurations of an ensemble at $T$ is about $10^{4}$.
The local $d$-wave order parameter now is evaluated as
$\la\Delta^{d}_{i}\ra=\frac{1}{4}\sum_{j}\la\Delta_{ij}\ra
\left[\delta_{j,i+{\hat x}}+\delta_{j,i-{\hat x}}-\delta_{j,i+{\hat y}}
-\delta_{j,i-{\hat y}}\right]$.
Following this procedure yields a statistical distribution $D$ 
of local order parameters $\Delta\equiv|\la\Delta^{d}_{i}\ra|$.
As explained earlier,
the probability $P(\Delta)$ is given by
$\partial_{\omega} D(\omega)|_{\Delta}$. In Fig.~1 we plot $P(\Delta)$ (in an arbitrary unit)
at various $T$ in the units of the hopping amplitude.
Other parameters used are $V=2.05$, $\mu=0$, and $J_{XY}=0.11$ and impurities are not considered.
At low $T$, $P(\Delta)$ shows a sharp peak at $\Delta=0.25$, which is the mean-field
value at $T=0$, because phase fluctuations are not significant. 
Even near $T_{c}\simeq \frac{1}{4}T_{MF}$,
$P(\Delta)$ is confined within $\Delta\simeq0.21\sim0.24$. 
As $T$ increases above $T_{c}$, the probability spreads; however, it sharpens back
when $T\rightarrow T_{MF}$ because the mean-field gap diminishes rapidly. 
Since the median of $P(\Delta)$ moves towards zero, the spatial average of the local
order parameters decreases. The inset of Fig.~1 demonstrates such a behavior
compared with its mean-field counterpart. In particular, the spatial average does not change
much until $T>1.5T_{c}$. This is consistent with a variation of the density of states
computed in Ref.\cite{eckl}. 
We also performed the same calculations
choosing $V=0.92$ for which $T_{c}\simeq \frac{5}{6}T_{MF}$ and obtained qualitatively
similar results.
Gomes {\it et al.}\cite{gomes}  measured local $d$-wave order parameters of \bsco
to show the order parameter distribution. They also presented the probability for
the overdoped \bsco
in a histogram with an interval of $2meV$, which is equivalent to $\epsilon$ in our definitions.
The histogram illustrates the probability spreads as $T$ increases while it becomes a sharp peak
at a high $T$. It is also indicated that the median of the probability moves towards zero
with increasing $T$.
The detailed behavior of the probability in Ref.~\cite{gomes}
is not completely identical to that shown in Fig.~1.
Nonetheless, important characteristics of the
histogram are congruent with Fig.~1. 

\begin{figure}[tp]\begin{center}
\includegraphics[height=2.5in,width=2.5in]{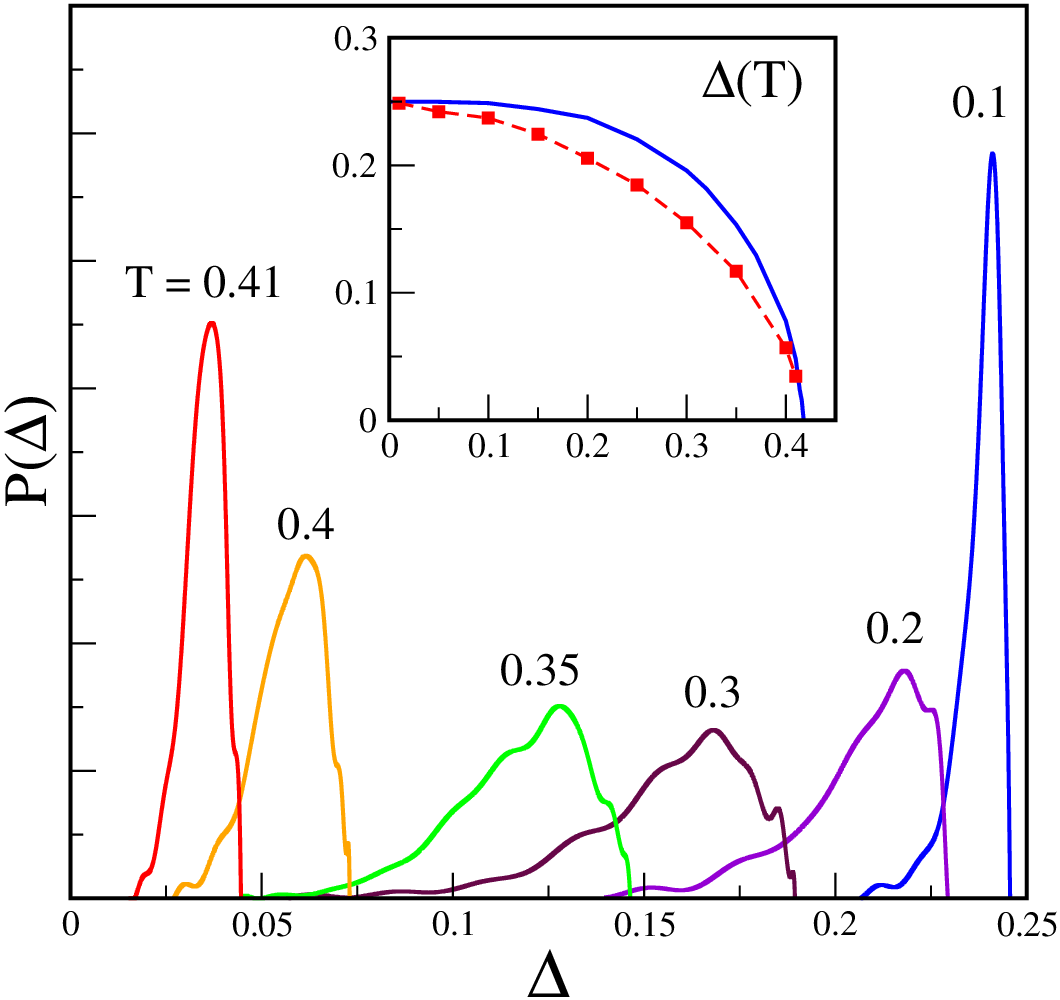}
\caption{(Color online)
Probability $P(\Delta)$ (in an arbitrary unit) to find the value of local
order parameters $\Delta$ at various $T$
in the unit of the hopping amplitude.
As $T$ increases, the probability distribution spreads.
However, $P(\Delta)$ shows a sharp distribution again when $T\rightarrow T_{MF}$.
The inset shows that the spatial average (symbol) of the local order parameters
diminishes as $T$ increases compared with the mean-field gap (curve).
}
\end{center}\end{figure}

Another interesting experimental observation using the STM 
has been recently reported by Chatterjee {\it et al.}\cite{chatterjee}.
They observed the evolution of the impurity resonance as $T$ increases
from low $T$ to high $T (>T_{c})$
and found that the resonance survives above $T_{c}$. Moreover, 
the resonance evolves smoothly with increasing $T$ and does not show any sudden 
broadening near $T_{c}$. In fact, it was argued that such a sudden broadening near $T_{c}$ 
would occur in the phase fluctuation scenario\cite{qwang}. This led Chatterjee {\it et al.}
to conclude that the phase fluctuation scenario is not consistent with their findings.
However, in Ref.\cite{qwang} the phase of the local order parameters is 
assumed to vary on the
length scale of the London penetration depth so that 
the quasiclassical approximation is applicable.
Within our framework, such an assumption corresponds to weak phase fluctuations 
and, thus, $T_{c}$ would not be much less than $T^{*}$. The only assumption we made following
Refs.~\cite{sharapov,eckl} is that
the phase varies slowly over the coherence length without any larger length scale employed,
and the BdG formalism is applied. 
Consequently, the phase varies at any length scale greater than the lattice constant,
and any degree of fluctuations can be considered.

The local differential conductance is described by the local
density of states (LDOS) in the theoretical calculations.
We apply the supercell technique to compute the LDOS as in Refs.\cite{zhu2,chen}.
If the lattice size is $N_{x}\times N_{y}$ as a unit cell and the number of unit cells is
$M_{x}\times M_{y}$, then the total size becomes $N_{x}M_{x}\times N_{y}M_{y}$.
We also introduce the quasimomenta $p_{x(y)}=\frac{2\pi n_{x(y)}}{N_{x(y)}M_{x(y)}}$,
where $n_{x(y)} = 0,1,\cdots M_{x(y)}-1$.
It is obvious that $\cZ_{BdG}$, now, depends on $p_{x(y)}$. In the presence of phase
fluctuations, one can think of two different supercell techniques depending on
how to associate the fluctuations with $p_{x(y)}$. One way is 
to start with the same random configuration of phases regardless of $p_{x(y)}$
while the other would be to use different random configurations for different values of
$p_{x(y)}$. Nonetheless between the two methods
we found no practical difference in the LDOS averaged over the ensemble.
In particular, for $T\geq T_{c}$ we obtain identical results.
We obtain the averaged LDOS as follows: the LDOS is calculated for a given phase
configuration and its average is computed over the ensemble.
For a single impurity, we choose a strong impurity potential $U=100$, 
which is close to the unitary limit\cite{salkola}. 
Other parameters are $V=1.1$, $\mu=0$, and $J_{XY}=0.033$. For the lattice size,
$N_{x}=N_{y}=24$ and $M_{x}=M_{y}=10$. In Fig.~2, we plot the LDOS at the nearest neighbor site to
the impurity (solid curves) as well as at a site far away from the impurity (dashed curves)
for $T=0.01$, $0.02$, $0.03$ and $0.04$.
Note that the impurity resonances survive above $T_{c}\approx0.03$.
Moreover, the resonance does not change much with increasing $T$ near $T_{c}$.
The displayed behavior is consistent with the experimental findings\cite{chatterjee}.
It is illustrated further in Fig.~3 for $V=1.1$ and for a weaker pairing potential 
$V=0.6$ at $T=0.02$, $0.03$, and $0.04$. This indicates that 
the impurity resonance above $T_{c}$ does not strongly depend on parameters.
From the theoretical point of view, it is originated from the fact that for these cases
phase fluctuations $(\theta_{ij})$ cannot disturb significantly the phase
difference of the mean-field value $\pi(={\bar\theta}_{i,i+\hat y}-{\bar\theta}_{i,i+\hat x})$.

{\it In conclusion,}
considering recent experiments on the imaging of the inhomogeneous
local order parameters above $T_{c}$
as well as the impurity resonance peak in high-$T_{c}$ superconductors, we calculated
a distribution of the local order parameters and the local density of states near an impurity
based on the phase fluctuation scenario.
The order parameter distribution is qualitatively
agreed with experimental observations. 
Randomly distributed impurities may be necessary for more realistic model calculations.
A smooth evolution of the impurity resonance peak, through $T_{c}$, seen in the STM
measurement also supports this picture. The resonance peak computed
within our framework does not show any sudden broadening near $T_{c}$ and consistent with
the experimental results. 

We acknowledge N.P. Ong for useful discussions. This work was supported by the Robert A. Welch
Foundation and the Texas Center for Superconductivity at the University of Houston through the
State of Texas, the Ministry of Science and Technology of China (973 project No: 2009CB929200),
NSFC Grant No. 10874032, and the Program for Professor of Special Appointment (Eastern Scholar) 
at Shanghai Institutions
of Higher Learning.

\begin{figure}[tp]\begin{center} 
\includegraphics[height=2.5in,width=2.5in]{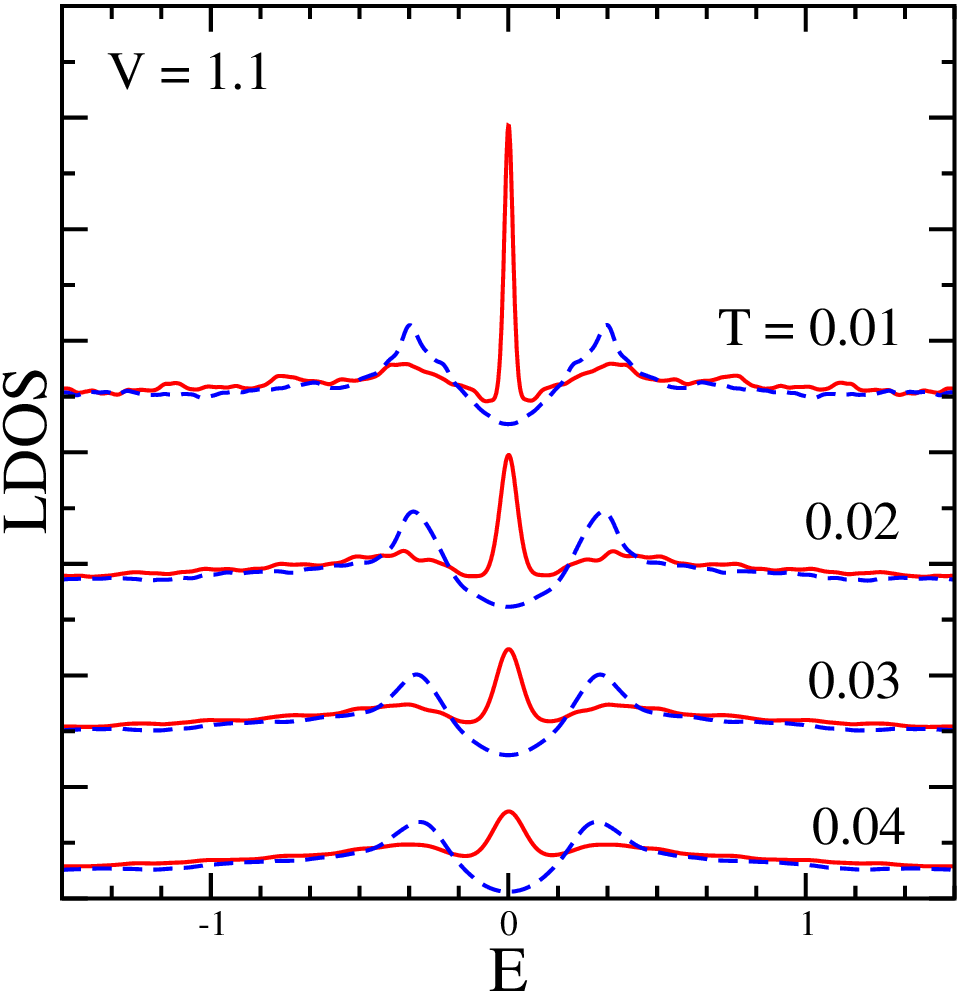}
\caption{(Color online)
LDOS at the nearest neighbor of
the impurity (solid curves) and far away from it (dashed curves)
for various $T$. Regardless of the value of $V$, the impurity resonance 
evolves smoothly with increasing $T$ and survives above $T_{c}\approx0.03$.
The relevant parameters are $V=1.1$, $U=100$, $\mu=0$, and $J_{XY}=0.033$.
} 
\end{center}\end{figure}

\begin{figure}[tp]\begin{center}
\includegraphics[height=2.5in,width=2.5in]{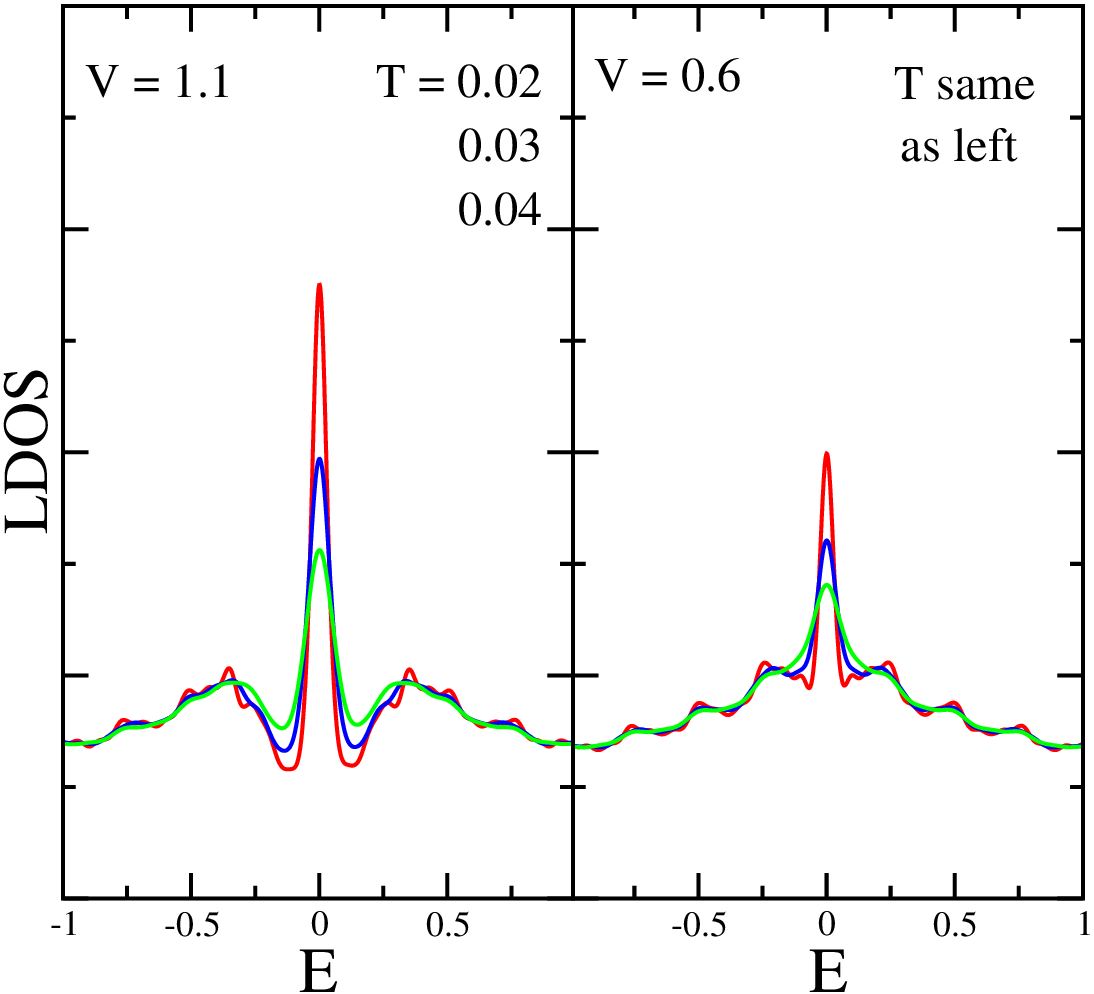}
\caption{(Color online)
LDOS at the nearest neighbor site of the impurity near $T_{c}$
for $V=1.1$ and $0.6$. Other parameters are same as in Fig.~2.
The resonance smoothly evolves with increasing $T$
from $0.02$ to $0.04$ through $T_{c}$. 
}
\end{center}\end{figure}

\bibliographystyle{prl}

\end{document}